\documentclass{ifacconf}

\usepackage{graphicx}      
\usepackage{natbib}        

\usepackage{amsmath}
\usepackage{amssymb}
\usepackage{enumerate}
\usepackage{enumitem}

\usepackage{xcolor}

\newcommand{\dom}{\operatorname{dom}}

\newif\ifitsdraft
\def\itsdraft{\global\itsdrafttrue}

\itsdraft  

\usepackage{xcolor}
\usepackage{placeins}

\usepackage{tikz}

\begin{document}
	\begin{frontmatter}
		
		\title{Contract-Based Design for Hybrid Dynamical Systems and Invariance Properties} 
		
		
		\author[First]{Sadek Belamfedel Alaoui} 
		\author[First]{Adnane Saoud} 
		
		\address[First]{College of Computing, University Mohammed VI Polytechnic, UM6P, Benguerir, Morocco\\ (e-mail: sadek.belamfedel, adnane.saoud (at) um6p.ma).}

		\begin{abstract}                
  \ifitsdraft
			This work establishes fundamental principles for verifying contract for interconnected hybrid systems. When system's hybrid arcs conform to the contract for a certain duration but subsequently violate it, the composition of hybrid dynamical systems becomes challenging. The objective of this work is to analyze the temporal satisfaction of the contract, allowing us to reason about the compositions that do not violate the contract up to a certain point in a hybrid time. Notions of weak and strong satisfaction of an assume-guarantee contract are introduced. These semantics permits the compositional reasoning on hybrid systems of varying complexity depending on the interconnection's type, feedback or cascade. The results show that both semantics are compatible with cascade composition, while strong semantic is required for feedback composition. Moreover, we have shown how one can go from weak to strong contract satisfaction. Finally, we have studied a particular class of hybrid systems and we have shown that the concept of forward (pre-)invariant relative to a contract makes it possible to deal with feedback compositions. These results are demonstrated throughout the paper with simple numerical examples. 
   \else
This work establishes principles for verifying contracts in interconnected hybrid systems. By analyzing temporal contract satisfaction, it enables reasoning about compositions that don't violate the contract up to a certain hybrid time point. Weak and strong satisfaction semantics are introduced for assume-guarantee contracts, facilitating compositional reasoning for different interconnection types. Under certain assumptions, it is possible to transition from weak to strong contract satisfaction. Additionally, forward (pre-)invariants relative to contracts aid in handling feedback compositions. Simple numerical examples illustrate these results.
   \fi
		\end{abstract}
		
		\begin{keyword}
			Assume-guarantee contracts; Hybrid dynamical systems; Forward invariance;
		\end{keyword}
		
	\end{frontmatter}
	
\section{Introduction}
\ifitsdraft
\else
\vspace{-10pt}
\fi

Modern Cyber-Physical Systems (CPS), which include autonomous vehicles, smart buildings, and robots, prominently feature hybrid systems as integral components. Verification techniques encounter limitations when dealing with complex CPS, mainly due to the significant number of interacting components. These limitations are compounded by the diverse behaviors exhibited by these components, which can include discrete jumps, continuous flows and hybrid behaviours, see \cite{saoud2018composition,shali2022composition,goebelsanfelice2012}. To address these challenges, the application of compositional approaches is crucial. These approaches, rooted in control theory, leverage principles such as the celebrated small-gain theorem, see \cite{jiang1994small}. Assume-Guarantee (AG) contract, a compositional method that has emerged from formal methods, offers valuable insights for the analysis and design of CPS, see \cite{alur2001compositional}.

AG contracts play a crucial role in the allocation of control goals among system components. Each contract defines an invariant, or more generally a temporal logic specification property that a component must satisfy, depending on certain assumptions about its interactions with the other components. While extensive work has laid the theoretical foundations for AG reasoning for dynamical systems and contract satisfaction in continuous and discrete time settings, as exemplified by studies such as \cite{kim2017small,saoud2018composition,girard2022invariant,di2020assume,shali2022composition} and related references, there are still gaps to be addressed. A significant challenge is reasoning about the composition of hybrid systems, which are known for their complex behavior, involving combinations of continuous flows and discrete jumps. Given the above room for contributing, this paper is motivated by two main goals. First, it aims to advance AG-style compositional verification techniques adapted to hybrid models. This motivation stems from fundamental questions, including how to formulate contracts for hybrid dynamical systems, how to establish criteria for contract satisfaction, and how to analyze satisfaction of local contracts of a composed system. Second, the relationship between AG contracts and the forward (pre-)invariance property in hybrid systems is investigated. This investigation shows how the existence of (pre-)invariants relative to contracts makes it possible to reason on feedback compositions.

%
%

The present work study the AG contracts for analyzing cascade and feedback compositions of hybrid dynamical systems. Cascade and feedback compositions are building blocks that allow to describe a wide range of interconnected systems. The starting point consists in considering the hybrid system as a set of hybrid arcs, thus allowing for all the possible behaviours (such as Zeno behaviors for example) and without requiring the hybrid arcs to be complete. This hybrid framework allows us to define the weak and strong semantics of AG contracts in order to allocate responsibilities among hybrid components. Then, the analysis of strong and weak contract satisfaction within the cascade and feedback composition is elaborated. Furthermore, the paper demonstrates a method for transitioning from weak contract satisfaction to the stronger one. Additionally, we show that forward (pre-)invariance property of a hybrid system leads to the weak satisfaction of a contract. Moreover, a reasoning for feedback composition for hybrid systems is established by relying on the concept of (pre-)invariants relative to contracts. Illustrative examples and figures are associated with the various results.  This paper focuses on the theoretical development of a general framework for contract-based reasoning for hybrid dynamical systems and invariance properties. Here, we emphasize that tools to verify whether a hybrid system satisfies a given contract, or to construct a controller to enforce the satisfaction of a contract, is beyond the scope of this paper and will be considered in future research.

\subsection{Related works}
\ifitsdraft
Contract based design have found applications in diverse fields, such as verifying aircraft electrical systems \cite{nuzzo2013contract}, synthesizing controllers in traffic networks \cite{kim2015compositional}, ensuring stability for embedded systems in the presence of timing uncertainty \cite{al2016verification}, and designing controllers for voltage stability and power sharing in time-varying DC microgrids \cite{zonetti2019symbolic}. 
\fi
Recent contributions extends the assume-guarantee contracts concept from the realm of formal methods to control systems, see e.g, \cite{kim2017small,saoud2021assume,shali2022composition}. For instance, the work in \cite{romeo2018quotient} proposed the operation of quotient for AG contracts, aiding compositional methodologies to address the missing component problem. The work in \cite{saoud2018composition} explores the composition of discrete and continuous-time dynamical systems using the notion contracts. The research in \cite{saoud2020contract} focuses on contract-based design for multiperiodic sampled-data systems. \cite{liu2022compositional} propose a compositional synthesis of signal temporal logic specifications through assume-guarantee contracts. Furthermore, the problem considered in \cite{girard2022invariant} involves assume-guarantee contracts with time-horizon characterization. The work in \cite{shali2022composition} develop tools to compare two contracts and identify the stricter one. These works have established theoretical foundations and efficient computational methods, but are limited to purely continuous or discrete time dynamical systems.

\ifitsdraft
In this paper, we also present a new approach to contract verification that differs from classical methods discussed in \cite{benveniste2018contracts}. Our approach introduces a progressive, temporal analysis of contract satisfaction that allows for the detection of contract violations at specific points in hybrid time, based on the hybrid time description in \cite{goebelsanfelice2012}. In \cite{benvenuti2014assume}, contracts are defined for hybrid automatas, for which solutions are solely parameterized by continuous-time. However, as it as been discussed in \cite{goebelsanfelice2012}, this formulation is insufficient for comprehensive modeling and analytical purposes. Regarding the contract-based approach in \cite{muller2016component}, a component is a combination of a discrete-time controller with a continuous-time system (described as a sampled-data system). In our approach we consider a more general setup where the system itself can be described as a hybrid system consisting of combinations of both discrete and continuous behaviors. 
\else
This paper introduces a novel contract verification approach, departing from classical methods \cite{benveniste2018contracts}. We propose a progressive, temporal analysis of contract satisfaction, enabling pinpointing of violations at specific points in hybrid time, following the framework by \cite{goebelsanfelice2012}. While contracts in \cite{benvenuti2014assume} are designed for hybrid automata, solely parameterized by continuous time, our approach, as discussed by \cite{goebelsanfelice2012}, provides a more comprehensive modeling and analytical framework. Furthermore, while \cite{muller2016component} focus on a component-based approach with discrete-time controllers and continuous-time systems, our method extends to a broader context, encompassing hybrid systems with both discrete and continuous behaviors.
\fi

\ifitsdraft
Finally, our approach differs from the one in \cite{benveniste2018contracts}. First, our approach introduces a progressive, temporal analysis of contract satisfaction that allows for the detection of contract violations at specific points in hybrid time, based on the hybrid time description in \cite{goebelsanfelice2012}. Second, while assumptions and guarantees cover all variables in \cite{benveniste2018contracts}, we focus on assumptions on inputs and guarantees on states and outputs. This deliberate restriction allows us to obtain simpler composition results. In addition, our method eliminates the need for saturated contracts, as it allows contracts to be handled directly without the need for union or complement computations, which is a requirement in the framework of \cite{benveniste2018contracts}.  
\fi

\subsection{Contributions and Organization}
Motivated by the absence of prior research on AG contracts and their relation to forward (pre-)invariance in hybrid systems, we introduce analytical tools for assessing AG contract compositions and analyse hybrid systems with forward (pre-)invariance properties.  Here, we emphasize that a fundamental challenge in formally composing hybrid systems is the disparity in the hybrid times of each component. Furthermore, this complexity is compounded by the potential for Zeno or incomplete behaviours. Section \ref{sec:2} outlines the fundamental framework of hybrid systems, encompassing inputs, states, outputs, and hybrid arcs. This formal representation paves the way for the introduction of formal concepts of weak and strong semantics in hybrid systems, as presented in Section \ref{sec:3}. We then present results on feedback and cascade composition of AG contracts with hybrid arcs. Indeed, we show that:
\begin{itemize}
	\item Both semantics are compatible with cascade composition (Theorems \ref{THM1} and \ref{THM2})
	\item Weak semantics are insufficient to reason about feedback composition. This is illustrated by Example \ref{Example1}.
	\item Strong semantics is required to reason about feedback composition (Theorem \ref{THM3}), supported by Example \ref{Example2};
	\item Under certain assumptions, weak contract satisfaction can lead to strong satisfaction (Proposition \ref{prop1}). 
\end{itemize}
In the Section \ref{sec:4}, we develop results for specific class of hybrid systems described by differential and difference inclusions \cite{goebelsanfelice2012}. Here, we have established:
\begin{itemize}
	\item Conditions to go from the forward (pre-)invariance property to the weak contract satisfaction (Proposition \ref{prop2});
	\item Forward (pre-)invariance relative to contracts is compatible with  feedback compositions (Theorem \ref{THM5}), complemented by the illustrative Example \ref{Example4}.
\end{itemize}

\ifitsdraft
\else
Due to space limitation, the proofs can be found in the online version of the paper available on this link\footnote{http://tinyurl.com/59cjucex}.
\vspace{-5pt}
\fi

\section{Notation and Preliminaries}
\label{sec:2}
\ifitsdraft
\else
\vspace{-10pt}
\fi

\paragraph*{Notation:} $\mathbb{Z}$, $\mathbb{N}$ and $\mathbb{N}_{> 0}$ denote the sets of integers, of non-negative integers, and of positive integers, respectively. $\mathbb{R}$,  $\mathbb{R}_{\geqslant0}$ and $\mathbb{R}_{>}$ denote the sets of real, of non-negative real and of positive real numbers, respectively.  Given a set-valued mapping $M: \mathbb{R}^{m} \rightrightarrows \mathbb{R}^{n}$, we denote the range of $M$ as $\operatorname{rge} M=\left\{y \in \mathbb{R}^{n}: y \in M(x), x \in\right.$ $\left.\mathbb{R}^{m}\right\}$ and the domain of $M$ as $\dom  M=\left\{x \in \mathbb{R}^{m}: M(x) \neq\right.$ $\emptyset\}$. For $x \in \mathbb{R}^{n},\|x\|$ denotes the Euclidean norm of $x$. For $\varepsilon \in \mathbb{R}_{\geqslant0}$ $A \subseteq \mathbb{R}^{n}$, the $\varepsilon$-expansion of $A$ is $\mathcal{B}_{\varepsilon}(A)=\left\{y \in \mathbb{R}^{n} \mid \exists x \in A,\|x-y\| \leq \varepsilon\right\}$. For $\varepsilon \in \mathbb{R}_{>}$ $A \subseteq \mathbb{R}^{n}$, the $\varepsilon$-contraction of $A$ is $\mathcal{B}_{-\varepsilon}(A)=\left\{y \in \mathbb{R}^{n} \mid \mathcal{B}_{\varepsilon}(y) \subseteq A \right\}$. We use $|x|_{K}$ to denote the distance from point $x$ to a closed set $K$, i.e., $|x|_{K}=\inf _{\xi \in K}\|x-\xi\|$. The closure of the set $K$ is denoted as $\overline{K}$. The set of boundary points of a set $K$ is denoted by $\partial K$ and the set of interior points of $K$ is denoted by $\operatorname{int}(K)$. 

\subsection{Hybrid systems}
To introduce the class of systems considered throughout this paper, we recall the notions of hybrid time domain and hybrid arcs, as in \cite{goebelsanfelice2012}.
\begin{defn}\label{DefHybridTime} \textit{(hybrid time domain)}
	A subset $E \subset \mathbb{R}_{\geq 0} \times \mathbb{N}$ is a compact hybrid time domain\footnote{$E$ is the collection of pairs comprising time intervals and their corresponding jump times.} if,
	$$
	E=\bigcup_{j=0}^{J-1}\left(\left[t_j, t_{j+1}\right], j\right),
	$$
	where $J\in \mathbb{N}$ and for some finite sequence of times $0=t_0 \leqslant t_1 \leqslant t_2 \leqslant \ldots \leqslant t_J$. It is a hybrid time domain if for all $(T, J) \in E, E \cap([0, T] \times\{0,1, \ldots, J\})$ is a compact hybrid domain.
	Furthermore,
	$$
	\begin{aligned}
		& \sup _t E=\sup \left\{t \in \mathbb{R}_{\geq 0}: \exists j \in \mathbb{N} \text { such that }(t, j) \in E\right\}, \\
		& \sup _j E=\sup \left\{j \in \mathbb{N}: \exists t \in \mathbb{R}_{\geq 0} \text { such that }(t, j) \in E\right\} .
	\end{aligned}
	$$
	with $\sup E=\left(\sup _t E, \sup _j E\right)$, and length $(E)=\sup_t E+\sup_j E$.
\end{defn}

\begin{defn}\textit{(hybrid arc)}
	A function $x: E \to \mathbb{R}^n$ is a hybrid arc if $E$ is a hybrid time domain and if for each $j \in \mathbb{N} \cup \{0, \ldots, \text{sup}_j E\}$, the function $t \mapsto x(t, j)$ is locally absolutely continuous on the interval $I^j=\{t:(t, j) \in E\}$.
\end{defn}

Building upon the preceding concepts, we define a hybrid system as follows:
\begin{defn}\label{Def4.3}
	A \textit{hybrid system} is a tuple $\mathcal{H}= (W, X, Y, \mathcal{T})$ where, 
	\begin{itemize}
		\item $W \subseteq \mathbb{R}^{m}, X \subseteq \mathbb{R}^{n}$ and $Y \subseteq \mathbb{R}^{p}$, are the sets of inputs, states, and outputs;
		\item $\mathcal{T}$ is a set of hybrid arcs $(w,x,y): \dom  x  \rightarrow$ $W \times X \times Y$ where $\dom  x   = \left\lbrace (t,j)\in \mathbb{R}_{\geq 0} \times \mathbb{N}: x(t,j) \neq \emptyset \right\rbrace$ with $\dom  x = \dom  w = \dom  y$. 
	\end{itemize}
	We say that the hybrid arc is \textit{complete} if $\dom  x$ is unbounded, i.e., if $\operatorname{length}(E) = \infty$.
\end{defn}


\begin{defn}
	A hybrid arc $(w,x,y): \dom x \to (W\times X\times Y) $ of a hybrid system $\mathcal{H}=(W, X, Y, \mathcal{T})$ is maximal if there does not exist another  hybrid arc $(w',x',y'): \dom x' \to (W\times X\times Y) \in \mathcal{T}$ such that $\dom  x $ is a proper subset of $\dom x  '$ and $x(t, j)=x'(t, j)$, $w(t,j)=w'(t,j)$, $y(t,j)=y'(t,j)$ for all $(t, j) \in \dom  x $.
\end{defn}

\ifitsdraft
The paper's results are formulated for the broader class of hybrid systems, as outlined in Definition \ref{Def4.3}. Subsequently, to illustrate the findings, we delve into the specific class of hybrid dynamical systems described by a combination of differential-difference inclusions. A more in-depth analysis of this particular case is presented in Section \ref{sec:4}.
\else
The results of the paper extend to a broader class of hybrid systems (Definition. \ref{Def4.3}). We then focus, for illustrative purposes, on hybrid dynamical systems combining differential-differential inclusions (Sec. \ref{sec:4}).
\fi

\begin{align}\label{Eq2}
	\overline{\mathrm{H}}: \begin{cases}
		(x,w) \in \mathrm C & \dot{x} \in \mathrm F(x,w) \\
		(x,w) \in \mathrm D & x^{+} \in \mathrm G(x,w)   \\
		x(0,0) \in X^0 &    y = h(x)
	\end{cases},
\end{align}
where $x\in \mathrm X$ is the state, $w\in \mathrm W$ is input, $y\in \mathrm Y$ is the output and $X^{0}\subseteq \mathrm X$ is the set of initial conditions, $C\subseteq \mathrm X\times \mathrm W$ is the flow set, $\mathrm D\subseteq X \times W$ is the jump set, $\mathrm W\subseteq \mathbb{R}^{m}$ is the input set, $\mathrm F: \mathrm C\times \mathrm W \rightrightarrows \mathbb{R}^{n}$ and $\mathrm G: \mathrm D\times \mathrm W\rightrightarrows \mathbb{R}^{n}$ are set valued maps. 

The hybrid dynamical system $\overline{\mathrm{H}}$ is a particular class of the hybrid dynamical system defined in Definition \ref{Def4.3}, represented by the tuple $\overline{\mathrm{H}} = (\mathrm{W}, \mathrm{X}, \mathrm{Y}, \mathrm{T})$. Here, $\mathrm{T}$ denotes the set of hybrid arcs $(w, x, y)$: $\dom x \rightarrow \mathrm{W} \times \mathrm{X} \times \mathrm{Y}$, solutions to $\overline{\mathrm{H}}$ if $\operatorname{dom} x=\operatorname{dom} y=\operatorname{dom} w, (x(0,0), w(0,0)) \in \overline{\mathrm{C}} \cup \mathrm{D}$, $y(t,j)=h(x(t,j))\in \mathrm{Y}$ for all $(t, j) \in \operatorname{dom} x$, and
\begin{itemize}
	\item[(S1)] for all $j \in \mathbb{N}$  such that $I^j$ has nonempty interior, and,
	$$
	\begin{aligned}
		\begin{cases}
			(x(t, j),w(t, j)) \in \mathrm{C} & \text{ for all } t\in \operatorname{int}(I^{j}), \\
			\dot{x}(t, j) \in \mathrm{F}(x(t, j), w(t, j))& \text{ for almost all } t\in I^{j},
		\end{cases}
	\end{aligned}
	$$
	\item[(S2)] for all $(t, j) \in \operatorname{dom} x$ such that $(t, j+1) \in \operatorname{dom} x$, $$(x(t, j),w(t, j))\! \in \! \mathrm{D}, \quad x(t, j+1)\! \in \! \mathrm{G}(x(t, j), w(t, j))\!$$

\end{itemize}

\subsection{Cascade and feedback compositions}

\ifitsdraft

Consider interconnections among hybrid systems that can be described using cascade or feedback compositions, as shown in Figure \ref{fig:composition}. Note that feedback and cascade compositions serve as fundamental building blocks for any interconnected system.

\begin{figure}[h]
	\centering
	\includegraphics[width=1\linewidth]{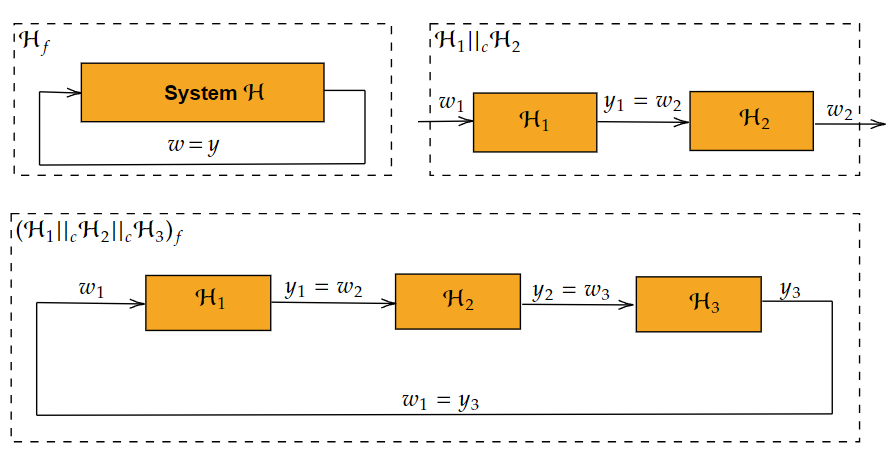}
	\caption{Cascade and feedback compositions of hybrid systems.}
	\label{fig:composition}
\end{figure}

\else
Consider interconnections among hybrid systems that can be described using cascade or feedback compositions.

\fi

\ifitsdraft
For the  cascade composition of two hybrid systems, as in Figure \ref{fig:composition}, the input of the forward system $w_{2}$ is a hybrid arc with the hybrid time domain $\dom x_{1}$, which does not necessarily match the one of the state hybrid arc $x_{2}$ defined by, $\dom x_{2}$. The composition of two hybrid dynamical systems $\mathcal{H}_{1}$ and $\mathcal{H}_{2}$ can be described along the smallest hybrid time domain $\dom x_{1}$ or $\dom x_{2}$. Furthermore, the solutions are described over a reparametrization of $\dom x_{1}$ and $\dom x_{2}$, as in \cite{Pauline2020}. This reparametrization aligns $\dom x_{1}$ and $\dom x_{2}$ to shared domain. The detailed procedure on the computation of the shared time domain is out of the scope of this paper and has been presented in \cite{Pauline2020}. An example of the construction of the shared time domain is provided below.

\begin{exmp}
	Consider two hybrid time domains $E_{1}$ and $E_{2}$ defined by, 
	$$\begin{aligned}
		E_1=&\left[0,1\right]\times \left\lbrace 0 \right\rbrace \cup \left[1,1.5\right]\times \left\lbrace 1 \right\rbrace  \\
		& \cup \left[1.5,1.5\right]\times \left\lbrace 2 \right\rbrace \cup \left[1.5,2\right]\times \left\lbrace 3 \right\rbrace,\\
		E_2=&\left[0,0.5\right]\times \left\lbrace 0 \right\rbrace \cup \left[0.5,2\right]\times \left\lbrace 1 \right\rbrace,
	\end{aligned}$$
	The shared hybrid domain $E_{12}$ is represented by,
	$$\begin{aligned}
		E_{12}=&\left[0,0.5\right]\times \left\lbrace 0 \right\rbrace \cup \left[0.5,1\right]\times \left\lbrace 1 \right\rbrace  \cup \left[1,1.5\right]\times \left\lbrace 2 \right\rbrace \\
		& \cup \left[1.5,1.5\right]\times \left\lbrace 3 \right\rbrace \cup \left[1.5,2\right]\times \left\lbrace 4 \right\rbrace,
	\end{aligned}$$
	Figure \ref{fig:domx1} illustrates $E_{1}$, $E_{2}$ and their shared hybrid time domain $E_{12}$.
\end{exmp} 	
\begin{figure}[!h]
		\begin{minipage}[t]{1\linewidth}
			\centering
			\includegraphics[width=0.9\linewidth]{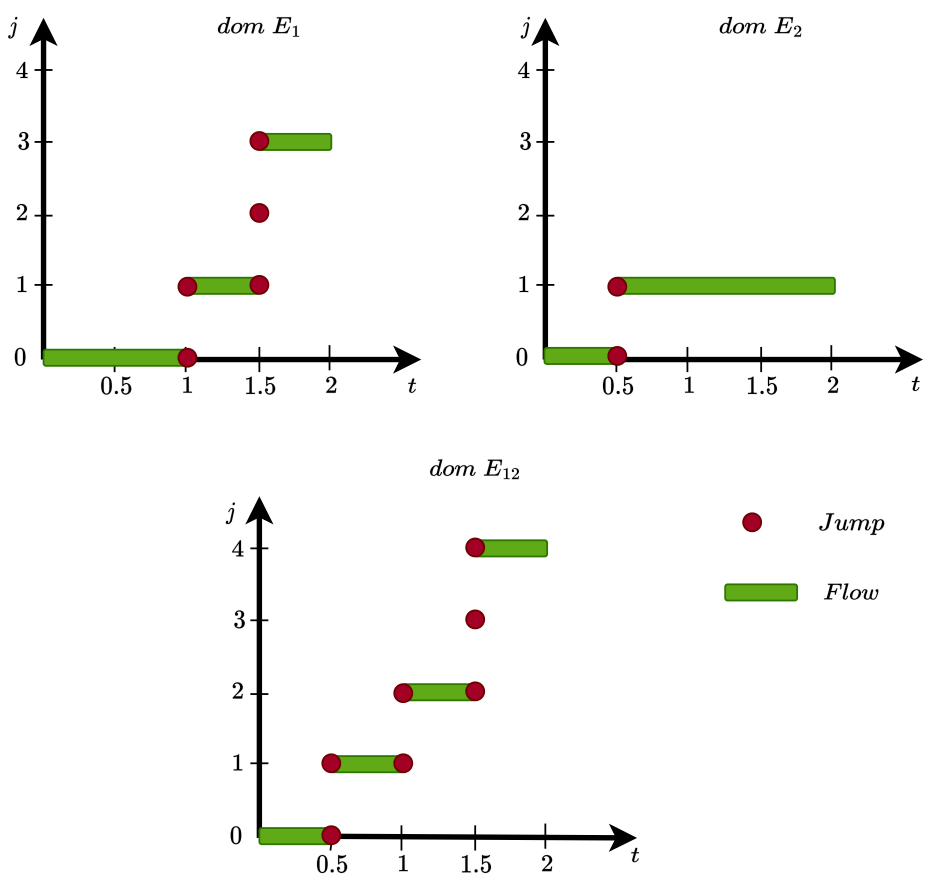}
			\caption{Graphical representation of $E_{1}$ and $E_{2}$ and their shared hybrid time domain $E_{12}$.}
			\label{fig:domx1}
		\end{minipage}  \\
		
	\end{figure} 

\else

For cascade composition, the output $y_{1}$ and input $w_{2}$ may not have the same hybrid time domain. Systems $\mathcal{H}_{1}$ and $\mathcal{H}_{2}$ can be composed over the smallest hybrid time $\dom x_{1}$ or $\dom x_{2}$, with solutions described over reparameterizations, as shown in \cite{Pauline2020}.

\fi

\begin{defn} (\textit{Cascade composition}) \label{Def-CascadeComposition}
	Consider the hybrid systems $\mathcal{H}_{1}=\left(W_{1}, X_{1}, Y_{1}, \mathcal{T}_{1}\right)$ and $\mathcal{H}_{2}=\left(W_{2}, X_{2}, Y_{2}, \mathcal{T}_{2}\right)$ with $Y_{1} \subseteq W_{2}$. The \textit{cascade composition} of $\mathcal{H}_{1}$ and $\mathcal{H}_{2}$ is the system $\mathcal{H}_{1} \|_{c} \mathcal{H}_{2}= \left(W_{1}, X_{1} \times X_{2}, Y_{2}, \mathcal{T}_{c}\right)$, such that $\left(w_{1},\left(x_{1}, x_{2}\right), y_{2}\right): \dom  x _{12} \rightarrow$ $W_{1} \times\left(X_{1} \times X_{2}\right) \times Y_{2}$ belongs to $\mathcal{T}_{c}$ if and only if there exist two hybrid arcs $\left(w_{1}, x_{1}, y_{1}\right): \dom  x _{12} \to W_{1} \times X_{1} \times Y_{1} $ in $\mathcal{T}_{1}$, and $\left(w_{2}, x_{2}, y_{2}\right): \dom  x _{12} \to W_{2} \times X_{2} \times Y_{2}$ in $\mathcal{T}_{2}$, with a shared time domain $\dom x_{12}$, and such that for all $(t,j) \in \dom  x _{12}, y_{1}(t,j)=w_{2}(t,j)$.
\end{defn}

\begin{defn}(\textit{Feedback composition}) \label{Def-feedbackComposition}
	Let $\mathcal{H}=(W, X, Y, \mathcal{T})$ be a hybrid system such that $Y \subseteq W$. The feedback composition of $\mathcal{H}$ is the system $\mathcal{H}_{f}=$ $\left(\{0\}, X,\{0\}, \mathcal{T}_{f}\right)$, such that $(0, x, 0): \dom  x  \rightarrow\{0\} \times X \times\{0\}$ belongs to $\mathcal{T}_{f}$ if and only if there exists $(w, x, y): \dom  x  \rightarrow$ $W \times X \times Y$ in $\mathcal{T}$ such that for all $(t,j) \in \dom  x , y(t,j)=w(t,j)$.
\end{defn}
Note that systems resulting from feedback composition have trivial null inputs and outputs. Hence, with an abuse of notation, we will denote $\mathcal{H}_{f}=\left(X, \mathcal{T}_{f}\right)$ and $x \in \mathcal{T}_{f}$, with $x: \dom  x  \to X$.

\section{AG contracts and compositional reasoning}
\label{sec:3}
\ifitsdraft
\else
\vspace{-10pt}
\fi

An AG contract is a compositional tool that specifies how a system behaves under assumptions about its inputs \cite{benveniste2018contracts}. The use of AG contracts makes it possible to reason on a global system based on properties of its components. This section aims first to define a contract along with its (weak$\backslash$strong) satisfaction for hybrid dynamical systems. The formal definition of an AG contract is as follows,
\begin{defn}\label{Def:contract}
	Let $\mathcal{H}=(W, X, Y, \mathcal{T})$ be a hybrid system, an AG contract for $\mathcal{H}$ is a tuple $\mathcal{C}=$ $\left(A_{W}, G_{X}, G_{Y}\right)$ where
	\begin{itemize}
		\item $A_{W} \subseteq W$ is a set of assumptions;
		\item $G_{X} \subseteq X$ and $G_{Y} \subseteq Y$ are sets of guarantees.
	\end{itemize}
\end{defn}

\ifitsdraft
\begin{figure*}
	\centering
	\includegraphics[width=15cm, height=9cm]{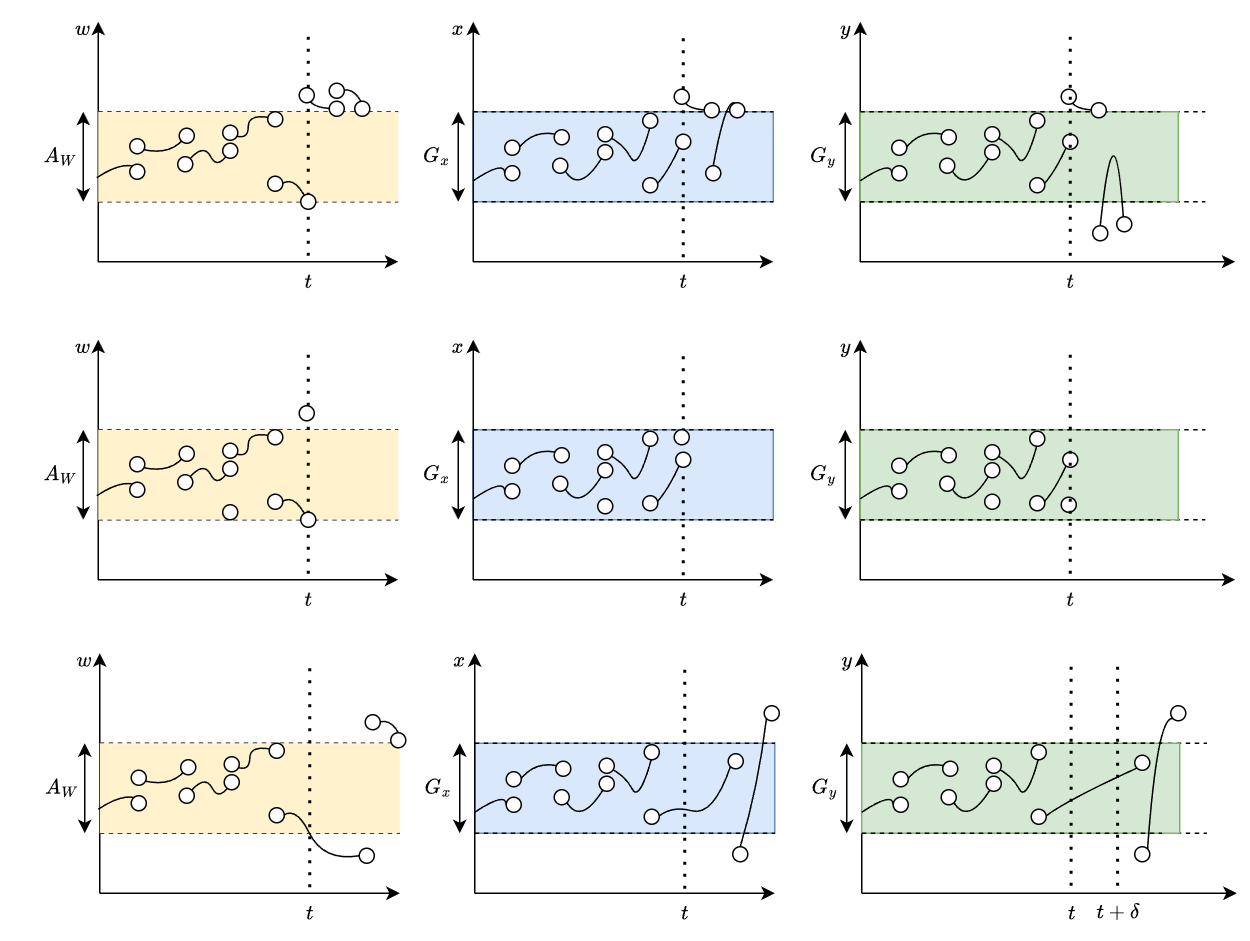}
	\caption{An illustration of weak and strong satisfaction of an AG contract. Top: Weak satisfaction; Middle: Strong satisfaction with jump; Bottom: Strong satisfaction with flow. Continuous curves: flow mode; Circles: jump mode.}
	\label{fig:WeakAndstrong}
\end{figure*}
\else

\fi

\subsection{Weak semantics}
In this subsection, we define the notion of (weak) AG-contract satisfaction for hybrid systems. The aim of this section is to show that (weak) AG contract satisfaction is sufficient for reasoning about cascade composition, whereas it is generally not sufficient for feedback composition. 

\begin{defn}
	\label{Def-ContractWeak}
	Let $\mathcal{H}=(W, X, Y, \mathcal{T})$ be a hybrid system and $\mathcal{C}=$ $\left(A_{W}, G_{X}, G_{Y}\right)$ be an AG contract.
	We say that $\mathcal{H}$ (weakly) satisfies $\mathcal{C}$, denoted $\mathcal{H} \models \mathcal{C}$, if for all hybrid arcs $(w, x, y): \dom  x  \to W \times X \times Y$ in $\mathcal{T}$, the following holds: for all $(T,J) \in \dom  x $, such that for all $(t,j) \in  \dom  x \cap([0, T] \times\{0,1, \ldots, J\}) $, $ w(t,j) \in A_{W}$, we have: 
		\begin{itemize}
			\item for all $ (t,j) \in  \dom  x \cap([0, T] \times\{0,1, \ldots, J\}), x(t,j) \in G_{X}$;
			\item for all $ (t,j) \in  \dom  x \cap([0, T] \times\{0,1, \ldots, J\}), y(t,j) \in G_{Y}$. 
		\end{itemize}
\end{defn}

\ifitsdraft
The core concept behind the (weak) satisfaction of a contract is that as long as  the assumptions are satisfied along a hybrid time domain the system's state and output are in their respective guarantee sets.   This condition is considered mild as it imposes no additional requirements prior to the hybrid time for the satisfaction of assumptions. The intuition for the (weak) semantics is depicted in Figure \ref{fig:WeakAndstrong} (Top graphs).
\fi


\subsubsection{Compositional reasoning}
The following results show that weak semantics can be used to reason about cascade composition.

\begin{thm}\label{THM1} (Contracts under cascade composition): Let $\mathcal{H}_{i}=\left(W_{i}, X_{i}, Y_{i}, \mathcal{T}_{i}\right), i=1,2$, be hybrid systems with $Y_{1} \subseteq W_{2}$. Let $\mathcal{C}_{i}=\left(A_{W_{i}}, G_{X_{i}}, G_{Y_{i}}\right)$ be assume-guarantee contracts for $\mathcal{H}_{i}, i=1,2$, with $G_{Y_{1}} \subseteq A_{W_{2}}$, and let $\mathcal{C}_{c}=\left(A_{W_{1}}, G_{X_{1}} \times G_{X_{2}}, G_{Y_{2}}\right)$. If $\mathcal{H}_{1} \models \mathcal{C}_{1}$ and $\mathcal{H}_{2} \models \mathcal{C}_{2}$, then $\mathcal{H}_{1} \|_{c} \mathcal{H}_{2} \models \mathcal{C}_{c}$.
\end{thm}

\ifitsdraft
\begin{pf}
	Let $\left(w_{1},\left(x_{1}, x_{2}\right), y_{2}\right): \dom  x_{12} \to W_{1} \times\left(X_{1} \times\right.$ $\left.X_{2}\right) \times Y_{2}$ in $\mathcal{T}_{c}$, where $\dom x_{12}$ is the shared hybrid time domain of $\dom x_{1}$ and $\dom x_{2}$. Then, there exist $\left(w_{1}, x_{1}, y_{1}\right): \dom  x_{12} \rightarrow$ $W_{1} \times X_{1} \times Y_{1}$ in $\mathcal{T}_{1}$ and $\left(w_{2}, x_{2}, y_{2}\right): \dom  x_{12} \to W_{2} \times X_{2} \times Y_{2}$ in $\mathcal{T}_{2}$ such that for all $(t,j) \in \dom  x_{12}, y_{1}(t,j)=w_{2}(t,j)$. 
	
	Let $(T,J) \in \dom  x_{12}$, such that for all $(t,j) \in \dom  x_{12} \cap([0, T]\times\{0,1, \ldots, J\}), w_{1}(t,j) \in A_{W_{1}}$. Satisfaction of $\mathcal{C}_{1}$ gives that for all $(t,j) \in \dom  x_{12} \cap ([0, T]\times\{0,1, \ldots, J\})$, $x_{1}(t,j) \in G_{X_{1}}$ and $y_{1}(t,j) \in G_{Y_{1}}$. Then, since $w_{2}(t,j)=y_{1}(t,j)$ we have for all $(t,j) \in \dom  x_{12} \cap([0, T]\times\{0,1, \ldots, J\}), w_{2}(t,j)=$ $y_{1}(t,j) \in G_{Y_{1}} \subseteq A_{W_{2}}$. Then, satisfaction of $\mathcal{C}_{2}$ gives that for all $(t,j) \in \dom  x_{12} \cap ([0, T]\times\{0,1, \ldots, J\}), x_{2}(t,j) \in G_{X_{2}}$ and $y_{2}(t,j) \in G_{Y_{2}}$. Hence, $\mathcal{H}_{1} \|_{c} \mathcal{H}_{2} \models \mathcal{C}_{c}$.
\end{pf}
\fi

The following counter-example shows that weak semantics is generally insufficient to reason on feedback composition.	

\begin{exmp} \label{ConterExample}\label{Example1}
	Let the system $\mathcal{H}_{1}=\left(W, X, Y, \mathcal{T}_{1}\right)$ where $W=X=Y=\mathbb{R}_{\geqslant 0}$. A hybrid arc of $\mathcal{H}_{1}$ is a triple $(w, x, y): \dom  x \to W \times X \times Y$ in $\mathcal{T}_{1}$ and where $\dom  x \subset \mathbb{R}_{\geq 0} \times \mathbb{N}$ and,
	$$
	\mathcal{H}_{1}: \begin{cases}x \in C & \dot{x}=F(x,w):=\sqrt[3]{w} \\ x \in D & x^{+}=G(x,w):=\frac{1}{2}w\\
		& y= x\end{cases}
	$$
	where the flow set is $C=\left\{x \in \mathbb{R}: |x| \leqslant 1\right\}$, the jump set is $D=\left\{x \in \mathbb{R}:|x| <2\right\}$ and $x(0,0)=0$. Obviously in the domain $C \cap D$ the state can evolve either with by flowing or jumping. Let us consider the AG contract $\mathcal{C}=$ $(\{0\},\{0\},\{0\})$ for $\mathcal{H}_{1}$. We can easily check that $\mathcal{H}_{1} \models \mathcal{C}$. However, the conclusion of the previous theorem does not hold for the feedback composition. Indeed, the hybrid arc defined by,
	$$x(t,j)= \left\lbrace \begin{array}{cc}
		(\frac{2}{3}t)^{\frac{3}{2}} & t\in \left[0,t_1\right], t_1=\big(\frac{3}{2}\big)^{\frac{2}{3}}, j=0 \\
		\frac{1}{2^{j}}	& t>t_1, j\geqslant 1 
	\end{array} \right., $$
	is a solution to the hybrid system $\mathcal{H}_f$ and there exists $(t,j) $ such that $ x(t,j) \notin G_{X}=\{0\}$.  
\end{exmp}

\ifitsdraft
From example \ref{ConterExample}, a crucial question arises: how can we formulate a new semantics that guarantees the compatibility of feedback composition? The following subsection introduces the concept of strong semantics for hybrid dynamical systems that allows effective reasoning about feedback composition.
\else
Example \ref{ConterExample} rises the question: How to ensure feedback composition? The next subsection presents strong semantics for hybrid systems, facilitating feedback composition.
\fi

\subsection{Strong semantics}

The aim of this part is to formulate a new semantics that guarantees the compatibility of feedback composition.

\begin{defn}\label{Def-Contract}
	Let $\mathcal{H}=(W, X, Y, \mathcal{T})$ be a hybrid system and $\mathcal{C}=$ $\left(A_{W}, G_{X}, G_{Y}\right)$ be an AG contract.
	We say that $\mathcal{H}$ strongly satisfies $\mathcal{C}$, denoted $\mathcal{H} \models_{s} \mathcal{C}$, if for all hybrid arcs $(w, x, y): \dom  x   \to W \times X \times Y$ in $\mathcal{T}$ :
	\begin{itemize}
		\item $y(0,0) \in G_{Y}$;
		\item for all $(T,J) \in \dom  x $, such that for all $ (t,j) \in  \dom  x \cap([0, T] \times\{0,1, \ldots, J\}),$ $ w(t,j) \in A_{W}$, the following conditions hold:
		\begin{itemize}
			\item[-] for all $ (t,j) \in  \dom  x \cap([0, T] \times\{0,1, \ldots, J\}), x(t,j) $ $ \in G_{X}$;
			\item[-] If $(T,J) \in \dom x$ is such that $(T,J+1) \in \dom x$, then for all $ (t,j) \in  \dom  x \cap([0, T] \times\{0,1, \ldots, J+1\}), y(t,j) \in G_{Y}$. Otherwise, if $(T,J) \in \dom x$ is such that $(T,J+1) \not\in \dom x$, there exists $\delta>0$ such that for all $ (t,j) \in  \dom  x \cap([0, T+\delta] \times \left\lbrace 0,1, \ldots, J \right\rbrace), y(t,j) \in G_{Y}$.
		\end{itemize}
	\end{itemize}
\end{defn}

\ifitsdraft
Let us remark that $\mathcal{H} \models_{s} \mathcal{C}$ obviously implies $\mathcal{H} \models \mathcal{C}$. Intuitively, the strong satisfaction of an AG contract for a hybrid system states that if the input of the system belongs to $A_{W}$ up to a hybrid time $(T,J) \in \dom  x $, then the state of the system belongs to $G_{X}$ at least until $(T,J)$ and the output of the system belongs to $G_{Y}$ at least until $(T+\delta,J)$ (if the systems is flowing after $(T,J)$) or $(T,J+1)$ (if the system jumps at $(T,J)$). The intuition for the (strong) semantics, whether in the jump or flow cases, is illustrated in Figure \ref{fig:WeakAndstrong} (middle and bottom graphs).

\else
Let us remark that $\mathcal{H} \models_{s} \mathcal{C}$ obviously implies $\mathcal{H} \models \mathcal{C}$.  It asserts that if the input belongs to $A_{W}$ until $(T,J) \in \dom x$, then the system's state adheres to $G_{X}$ at least until $(T,J)$. Moreover, the output belongs to $G_{Y}$ until $(T+\delta,J)$ or $(T,J+1)$, depending on whether the system flows or jumps after $(T,J)$.
\fi

\subsubsection{Compositional reasoning}

We now demonstrates that strong semantics permit to reason on cascade and feedback compositions. We begin first by presenting the reasoning under cascade composition of two hybrid systems:

\begin{thm}\label{THM2} (Contracts under cascade composition): Let $\mathcal{H}_{i}=\left(W_{i}, X_{i}, Y_{i}, \mathcal{T}_{i}\right), i=1,2$, be hybrid systems with $Y_{1} \subseteq W_{2}$. Let $\mathcal{C}_{i}=\left(A_{W_{i}}, G_{X_{i}}, \right. $ $ \left. G_{Y_{i}}\right)$ be AG contracts for $\mathcal{H}_{i}, i=1,2$ with $G_{Y_{1}} \subseteq A_{W_{2}}$, and let $\mathcal{C}_{c}=\left(A_{W_{1}}, G_{X_{1}} \times G_{X_{2}}, G_{Y_{2}}\right)$. The following implications hold:
	\begin{itemize}
		\item If $\mathcal{H}_{1}\models_{s} \mathcal{C}_{1}$ and $\mathcal{H}_{2}\models\mathcal{C}_{2}$, then $\mathcal{H}_{1} \|_{c} \mathcal{H}_{2} \models_{s} \mathcal{C}_{c}$;
		\item If $\mathcal{H}_{1} \models \mathcal{C}_{1}$ and $\mathcal{H}_{2} \models_{s} \mathcal{C}_{2}$, then $\mathcal{H}_{1} \|{ }_{c} \mathcal{H}_{2} \models_{s} \mathcal{C}_{c}$.
	\end{itemize}
\end{thm}

\ifitsdraft
\begin{pf}
	Let $\left(w_{1},\left(x_{1}, x_{2}\right), y_{2}\right): \dom  x_{12} \to W_{1} \times\left(X_{1} \times\right.$ $\left.X_{2}\right) \times Y_{2}$ in $\mathcal{T}_{c}$. Then, there exists $\left(w_{1}, x_{1}, y_{1}\right): \dom  x_{12} \rightarrow$ $W_{1} \times X_{1} \times Y_{1}$ in $\mathcal{T}_{1}$ and $\left(w_{2}, x_{2}, y_{2}\right): \dom  x_{12} \to W_{2} \times X_{2} \times Y_{2}$ in $\mathcal{T}_{2}$ such that for all $(t,j) \in \dom  x_{12}, y_{1}(t,j)=w_{2}(t,j)$. 
	
	Implication 1 - Let $\mathcal{H}_{1} \models_{s} \mathcal{C}_{1}$ and $\mathcal{H}_{2} \models \mathcal{C}_{2}$. Strong satisfaction of $\mathcal{C}_{1}$ gives that $y_{1}(0,0) \in G_{Y_{1}}$. Then, since $w_{2}(t,j)=y_{1}(t,j)$ for all $(t,j) \in \dom  x_{12} \cap([0, T]\times\{0,1, \ldots, J\})$, we have $w_{2}(0,0)=y_{1}(0,0) \in G_{Y_{1}} \subseteq A_{W_{2}}$. Then, satisfaction of $\mathcal{C}_{2}$ gives that $y_{2}(0,0) \in G_{Y_{2}}$. Now, let $(T,J) \in \dom  x_{12}$, such that for all $(t,j) \in \dom  x_{12} \cap([0, T]\times\{0,1, \ldots, J\}), w_{1}(t,j) \in A_{W_{1}}$. Strong satisfaction of $\mathcal{C}_{1}$ gives that for all $(t,j) \in \dom  x_{12} \cap([0, T]\times\{0,1, \ldots, J\}), x_{1}(t,j) \in G_{X_{1}}$  and one of the following two cases: 
	\begin{itemize}
		\item[(a)] $(T,J+1) \in \dom x_{12}$, and for all $ (t,j) \in  \dom  x_{12}\cap([0, T] \times\{0,1, \ldots, J+1\}), y_{1}(t,j) \in G_{Y_{1}}$.
		\item[(b)] $(T,J+1) \notin \dom x_{12}$ and there exists $\delta>0$ such that for all $ (t,j) \in  \dom  x_{12} \cap([0, T+\delta] \times \left\lbrace 0,1, \ldots, J \right\rbrace), y_{1}(t,j) \in G_{Y_{1}}$.
	\end{itemize}
	Then, since $w_{2}=y_{1}$ we have for all $(t,j) \in \dom  x_{12} \cap([0, T]\times\{0,1, \ldots, J\})$, $w_{2}(t,j)=y_{1}(t,j) \in G_{Y_{1}} \subseteq A_{W_{2}}$. Then, satisfaction of $\mathcal{C}_{2}$ implies that we have statement (c) or (d). 
	\begin{itemize}
		\item[(c)]  $(T,J+1) \in \dom x_{12}$, and for all $ (t,j) \in  \dom  x_{12}\cap([0, T] \times\{0,1, \ldots, J+1\}), y_{2}(t,j) \in G_{Y_{2}}$.
		\item[(d)] $(T,J+1) \notin \dom x_{12}$ and there exists $\delta>0$ such that for all $ (t,j) \in  \dom  x_{12} \cap([0, T+\delta] \times \left\lbrace 0,1, \ldots, J \right\rbrace), y_{2}(t,j) \in G_{Y_{2}}$.
	\end{itemize}
	Hence, $\mathcal{H}_{1} \|_{c} \mathcal{H}_{2}\models_{s} \mathcal{C}_{c}$.
	
	Implication 2 - Let $\mathcal{H}_{1} \models \mathcal{C}_{1}$ and $\mathcal{H}_{2} \models_{s} \mathcal{C}_{2}$. Strong satisfaction of $\mathcal{C}_{2}$ gives that $y_{2}(0,0) \in G_{Y_{2}}$. Now, let $(T,J) \in \dom  x_{12}$, such that for all $(t,j) \in \dom  x_{12} \cap ([0, T]\times\{0,1, \ldots, J\}), w_{1}(t,j) \in A_{W_{1}}$. Satisfaction of $\mathcal{C}_{1}$ gives that for all $(t,j) \in \dom  x_{12} \cap ([0, T]\times\{0,1, \ldots, J\}), x_{1}(t,j) \in G_{X_{1}}$ and $y_{1}(t,j) \in G_{Y_{1}}$. In addition, since for all $(t,j) \in \dom x_{12}\cap( [0, T]\times\left\lbrace 0, \dots, J\right\rbrace)$ $w_{2}(t,j)=y_{1}(t,j)$, we have $w_{2}(t,j)=y_{1}(t,j) \in G_{Y_{1}} \subseteq A_{W_{2}}$. Thus, strong satisfaction of $\mathcal{C}_{2}$ gives that for all $(t,j) \in \dom x_{12}\cap([0, T]\times\{0,1, \ldots, J\}), x_{2}(t,j) \in G_{X_{2}}$ and either statement (c) or (d) is satisfied. Hence, $\mathcal{H}_{1} \|_{c} \mathcal{H}_{2} \models_{s} \mathcal{C}_{c}$.
\end{pf}
\fi

For feedback composition, the guarantee set $G_{Y}$ is required to be closed. The result is as follows:

\begin{thm}\label{THM3}	(Contracts under feedback composition): Let $\mathcal{H}=(W, X, Y, \mathcal{T})$ be a hybrid system with $Y \subseteq W$ and let $\mathcal{H}_{f}=\left(X, \mathcal{T}_{f}\right)$ be the composed feedback system. Let $\mathcal{C}=\left(A_{W}, G_{X}, G_{Y}\right)$ be an assume guarantee contract for $\mathcal{H}$ with $G_{Y} \subseteq A_{W}$ and with a closed set of guarantees on the output $G_{Y}$. If $\mathcal{H} \models_{s} \mathcal{C}$ then, for all hybrid arcs $x: \dom  x  \to X$ in $\mathcal{T}_{f}$, we have for all $(t,j) \in \dom  x $, $x(t,j) \in G_{X}$.
\end{thm}

\ifitsdraft
\begin{pf}
	To show the result, it is enough to demonstrate it for maximal hybrid arcs of $\mathcal{T}_{f}$. Now, let $x$ : $\dom  x  \to X$ be a maximal hybrid arc in $\mathcal{T}_{f}$, then there exists $(w, x, y): \dom  x   \to W \times X \times Y$ in $\mathcal{T}$ such that $w(t,j)=y(t,j)$ for all $(t,j) \in \dom x$. Let us define, 
	\begin{align}\label{Eq1}
		(T,J)= \sup\left\lbrace (t,j) \in \dom  x  \mid  y(t,j) \in G_{Y} \right\rbrace.
	\end{align}
	Strong satisfaction of $\mathcal{C}$ gives that $y(0,0) \in G_{Y}$, hence $(T,J)$ is well defined.
	Let us show that $ T=\sup_{t} \dom x $ and $ J=\sup_{j} \dom x $. Proceeding by contradiction, we have the following cases:
	\begin{itemize}
		\item \underline{ Case 1: $T<\sup_{t} \dom x$ and $ J= \sup_{j} \dom x $.} Using the absolute continuity of $y$ during flows intervals and since $G_{Y}$ is closed, we have that for all $(t,j) \in \dom x \cap ([0, T] \times  \left\lbrace 0,\dots,J\right\rbrace), y(t,j) \in G_{Y}$. Since $w = y$ and $G_{Y}\subseteq A_{W}$, we also have for all $(t,j)\in \dom x \cap( \left[0,T\right]\times\left\lbrace 0,\dots,J\right\rbrace)$, $w(t,j) \in A _{W}$. Since the hybrid arc can only evolve by flowing, strong satisfaction of $\mathcal{C}$ gives that there exists $\delta \in \left(0,\sup_{t} \dom x-T\right] $ such that for all $(t,j) \in\dom x \cap([0, T+\delta] \times  \left\lbrace 0,\dots,J\right\rbrace), y(t,j) \in G_{Y}$. This contradicts the definition of $T$ given by \eqref{Eq1}, which shows that this case is actually impossible.
		\item  \underline{Case 2: $T=\sup_{t} \dom x$ and $ J< \sup_{j} \dom x $} Using the absolute continuity of $y$ during flows intervals and since $G_{Y}$ is closed, we have that for all $(t,j) \in \dom x \cap ([0, T] \times  \left\lbrace 0,\dots,J\right\rbrace), y(t,j) \in G_{Y}$. Since $w = y$ and $G_{Y}\subseteq A_{W}$, we also have for all $(t,j)\in \dom x \cup (\left[0,T\right]\times\left\lbrace 0,\dots,J\right\rbrace)$, $w(t,j) \in A _{W}$. Since the hybrid arc can only evolve by jumping, strong satisfaction of $\mathcal{C}$ gives that $(T,J+1)\in \dom x$ and that for all $(t,j) \in \dom x \cap ([0, T] \times  \left\lbrace 0,\dots,J+1\right\rbrace), y(t,j) \in G_{Y}$. This contradicts the definition of $J$ given by \eqref{Eq1}, which shows that this case is actually impossible.
		\item  \underline{Case 3: $T<\sup_{t} \dom x$ and $ J< \sup_{j} \dom x $.} 
		The contradiction in this case can be easily shown by combining the arguments of case $1$ and case $2$ above.
	\end{itemize}
	Hence, the only case that is possible is $T=\sup_{t} \dom x $ and $J=\sup_{j} \dom x $, and one has that for all $(t,j) \in \dom x$, $y(t,j)\in G_{Y}$. Then, since $w=y$ we have $w(t,j)=y(t,j) \in G_{Y} \subseteq A_{W}$ or all $(t,j) \in \dom x$. Then, strong satisfaction of $\mathcal{C}$ gives that $x(t,j) \in G_{X}$ for all $(t,j) \in \dom x$, which ends the proof.
\end{pf}
\fi

To illustrate the obtained findings, we present a modified scenario building upon the previous example, which result in the strong satisfaction of the contract.
\begin{exmp} \label{Example2}
	Let the system $\mathcal{H}_{2}=\left(W, X, Y, \mathcal{T}_{2}\right)$ where $W=X=Y=\mathbb{R}_{\geqslant 0}$. A hybrid arc of $\mathcal{H}_{2}$ is a triple $(w, x, y): \dom  x \to W \times X \times Y$ in $\mathcal{T}_{2}$ where $\dom  x \subset \mathbb{R}_{\geq 0} \times \mathbb{N}$ and,
	$$
	\resizebox{1\hsize}{!}{$\mathcal{H}_{2}: \begin{cases}
			x \in C & \; \dot{x} \;= F(x,w):=\sqrt[3]{w}  \\ 
			x \in D & x^{+}=G(x,w):= \frac{1}{2}w   \\ 
			t_j < t_{j+1} \text{ and } t_{j} \leqslant t\leqslant t_{j}+\frac{t_{j+1}-t_{j}}{2} < t_{j+1} & y= 0   ,   \\
		 t_j < t_{j+1} \text{ and } \frac{t_{j+1}-t_{j}}{2} < t < t_{j+1} & y= x,\\
			 t_j = t_{j+1}     & y= x,       
		\end{cases}$}
	$$
	where the flow set is $C=\left\{x \in \mathbb{R}: |x| \leqslant 1\right\}$, the jump set is $D=\left\{x \in \mathbb{R}:|x| <2\right\}$, $\dom x=[t_0,t_1]\times \{0\} \cup [t_1,t_2]\times \{1\} \cup [t_2,t_3]\times \{3\} \ldots $ and $x(0,0)=0$. Let us consider the AG contract $\mathcal{C}=$ $(\{0\},\{0\},\{0\})$ for $\mathcal{H}_{2}$. We can easily check that $\mathcal{H}_{2} \models_{s} \mathcal{C}$, where the value of $\delta$ as in Definition \ref{Def-Contract} is given for $(t,j) \in \dom x$ with $t \in [t_j,t_{j+1}]$ by $\delta=\frac{t_{j+1}-t_{j}}{2}$. We can also check that the conclusion of the previous theorem holds since the only hybrid arc $x: \dom  x \to X$ in $\mathcal{T}_{f}$ is given by $x(t,j)=0$ for all $(t,j) \in \dom x$. 
\end{exmp}

The importance of a strong semantics is now obvious, as it enables reasoning about the feedback composition. Thus, the following research question is important: what are the conditions on the system and contracts to deal with feedback composition without the need of strong contract satisfaction?

\subsection{From weak to strong contract satisfaction}




In this section, we will show that, with specific additional assumptions, it becomes feasible to transition from the weak satisfaction of AG contracts to the strong satisfaction by relaxing the set of guarantees, and thus enabling the analysis of feedback compositions.

\begin{prop}\label{prop1}
	Let $\mathcal{H}=(W, X, Y, \mathcal{T})$ be a hybrid system and let $\mathcal{C}=\left(A_{W}, G_{X}, G_{Y}\right)$ be an AG contract for $\mathcal{H}$. Let us assume that for all hybrid arcs $(w, x, y):\dom x \to W\times X \times Y \in \mathcal{T}$, we have the existence of $\beta \geqslant 0$ such that for all $(t,j) \in \dom  x  $ with $(t,j+1) \in \dom  x$, we have bounded variation of the output $y$ during jumps with $\left|y(t,j) - y(t,j+1)\right|  \leqslant \beta$. If $\mathcal{H} \models \mathcal{C}$, then, for all $\varepsilon>\beta$, $\mathcal{H} \models_{s} \mathcal{C}_{\varepsilon}$ where $\mathcal{C}_{\varepsilon}=$ $\left(A_{W}, G_{X}, \mathcal{B}_{\varepsilon}\left(G_{Y}\right) \cap Y\right)$.
\end{prop}

\ifitsdraft
\begin{pf}
	 Let us consider a positive scalar $\varepsilon$ satisfying $\varepsilon > \beta \geqslant 0 $ and let $\left(w, x, y\right): \dom  x  \to W \times X \times Y \in \mathcal{T}$, then $y(0,0) \in G_Y \subset \mathcal{B}_{\varepsilon}\left(G_Y\right) \cap Y$. Let $ (T,J) \in \dom  x $ such that for all $(t,j) \in \dom  x \cap (\left[0,T\right], \left\lbrace 0, \dots, J \right\rbrace)$, $w(t,j) \in A_{W}$. Then, satisfaction of $\mathcal{C}$ implies that for all $ (t,j) \in \dom  x \cap  (\left[0,T\right]\times\left\lbrace 0,\dots,J\right\rbrace) $, $x(t,j) \in G_{X}$ and $y(t,j) \in G_{Y}$. To complete the proof, we consider two distinct cases: 
	\begin{itemize}
		\item \underline{Case 1:} If $(T,J) \in \dom x$ is such that $(T,J+1) \in \dom x$, then it follows from the bounded variation of the output during jumps that $ \left| y(T,J) - y(T,J+1)\right| \leqslant\beta $. Hence, one gets that for all $(t,j) \in \dom  x \cap  (\left[0,T\right]\times\left\lbrace 0,\dots,J+1\right\rbrace) $,  $y(t,j) \in \mathcal{B}_{\beta}\left(G_Y\right)\cap Y \subset \mathcal{B}_{\varepsilon}\left(G_Y\right)\cap Y$. 
		\item \underline{Case 2:} If $(T,J) \in \dom x$ is such that $(T,J+1) \notin \dom x$, it follows from the absolute continuity of $y$ that for all $(t,j) \in \dom x \cap  (\left[0,T\right]\times\left\lbrace 0,\dots,J+1\right\rbrace)$, there exists $\delta \neq 0 $ such that for all $(t,j) \!\in \!\dom  x \; \cap([T,T+\delta]\times\left\lbrace 0,\dots,J\right\rbrace)$, $y(t,j) \in \mathcal{B}_{\varepsilon}(G_{Y}) \cap Y$. Hence, one gets that for all $(t,j) \!\in \!\dom  x \; \cap( [0,T+\delta]\times\left\lbrace 0,\dots,J\right\rbrace)$, $y(t,j) \in \mathcal{B}_{\beta}\left(G_Y\right)\cap Y \subset \mathcal{B}_{\varepsilon}(G_{Y}) \cap Y$.
	\end{itemize}
\end{pf}
\fi

The following result on feedback composition, stated without proof, is a direct consequence of Proposition \ref{prop1} and Theorem \ref{THM3}:

\begin{cor}\label{cor1}
	Let $\mathcal{H}=(W, X, Y, \mathcal{T})$ be a hybrid system with $Y \subseteq W$ and let $\mathcal{H}_{f}=\left(X, \mathcal{T}_{f}\right)$ be its feedback composition. Let us assume that for all hybrid arcs $(w, x, y):\dom x \to W\times X \times Y \in \mathcal{T}$, we have the existence of $\beta \geqslant 0$ such that for all $(t,j) \in \dom  x  $ with $(t,j+1) \in \dom  x$, we have bounded variation of the output $y$ during jumps with $\left|y(t,j) - y(t,j+1)\right|  \leqslant \beta$. Let $\mathcal{C}=\left(A_{W}, G_{X}, G_{Y}\right)$ be an AG contract for $\mathcal{H}$ and let us assume the existence of an $\varepsilon > \beta$ such that $\mathcal{B}_{\varepsilon}\left(G_{Y}\right) \cap Y \subseteq$ $A_{W}$. 
	If $\mathcal{H}\models \mathcal{C}$, then, for all hybrid arcs $x: \dom  x  \to X $ in $\mathcal{T}_{f}$, we have for all $(t,j) \in \dom  x$, $ x(t,j) \in G_{X}$.
\end{cor}


\begin{exmp}\label{Example3}
	Consider the hybrid system $\mathcal{H}_{3}=(W, X, Y, \mathcal{T})$ where $W=X=Y=\mathbb{R}_{\geqslant 0}$. A hybrid arc of $\mathcal{H}_{3}$ is a triple $(w, x, y): \dom  x \to W \times X \times Y$ in $\mathcal{T}$ where for all $(t,j) \in \dom  x$,
	$$
	\mathcal{H}_{3}:\begin{cases}
		x \in C & \dot{x} =\sqrt{w}-x , \\
		x \in D & x^{+} = 0.1w   ,    \\
		& y = x                 
	\end{cases}
	$$
	where the flow set is $C=\left\{x \in \mathbb{R}: |x| < 0.9\right\}$, the jump set is $D=\left\{x \in \mathbb{R}:|x| \geqslant0.9\right\}$ and $x(0,0)=0.1$. Let $a>1$, let us consider the AG contract $\mathcal{C}=\left(A_W, G_X, G_Y\right)$ for $\mathcal{H}_{3}$ where $A_W=\left[0, a^2\right]$, $G_X= G_Y=[0, a]$, with $a \geq 11$. It is easy to check that $\mathcal{H}_{3} \models \mathcal{C}$ and that for all hybrid arcs $(w, x, y) \in \mathcal{T}$, we have the existence of an  $\beta=10a$ such that $\left| y^{+}-y\right|\leqslant \beta $. Moreover, for $\varepsilon= \beta=10a$, we have $\mathcal{B}_{\varepsilon}\left(G_Y\right) \cap Y=[0, 11a] \subseteq\left[0, a^2\right]=A_W$, since $a \geq 11$. Then, from corollary \ref{cor1}, it follows that for all hybrid arcs $x : \dom x \to X$ in $\mathcal{T}_{f}$, $x(t,j) \in \left[0, a\right]$ for all $ (t,j) \in \dom x$. Here, we emphasize that there exists non-zero hybrid arcs in $\mathcal{T}_{f}$ such as $x: \dom x  \to X$. Indeed, the hybrid arc  defined by, 
	$$
	x(t,j)= \left\lbrace \begin{array}{cc}
		\left(1-e^{-t / 2}\right)^{2} & t\in \left[0,t_1\right], t_1=5.943,j=0 \\
		\frac{1}{10^j}& t=t_1, j=1
	\end{array} \right.,
	$$
	is a solution to the hybrid system $\mathcal{H}_f$.
\end{exmp}

It is important to emphasise that Corollary \ref{cor1} does not hold when $\mathcal{B}_{\varepsilon}\left(G_{Y}\right) \cap Y \not\subseteq A_{W}$, for any $\varepsilon> \beta$, and in particular when $G_{Y}=A_{W}$. In such scenarios, how to reason about feedback composition based on weak semantics becomes challenging. One key property that is useful in this context is the notion of invariance in hybrid dynamical systems, as developed in \cite{chai2018forward}. To explore this challenge in more detail and to explore possible solutions, we turn to the following section.

\section{COMPOSITIONAL (pre-)iNVARIANTS FOR HYBRID INCLUSIONS}
\label{sec:4}
\ifitsdraft
\else
\vspace{-10pt}
\fi


In this section, our attention is directed towards hybrid systems denoted as $\mathcal{H}=\left(W, X, Y, \mathcal{T}\right)$, characterized by differential inclusions such as those expressed in \eqref{Eq2}. To begin, we revisit the notion of forward (pre-)invariance in the context of hybrid dynamical systems from \cite{chai2018forward} and the notion tangent cone from \cite{aubin2009set}.

\begin{defn}Forward (pre-)invariance of hybrid dynamical systems
	\begin{itemize}
		\item A set $K \subset \mathbb{R}^{n}$ is said to be forward (pre-)invariant for $\mathcal{H}$, as in \eqref{Eq2}, if every solution $(w,x,y): \dom  x  \rightarrow$ $W \times X \times Y$ for $\mathcal{H}$ satisfying $x(0,0) \in K$ we have that $ x(t,j) \in K$ for all $(t,j) \in \dom x$.
		\item A set $K \subset \mathbb{R}^{n}$ is said to be forward invariant for $\mathcal{H}$, as in \eqref{Eq2}, if every solution $(w,x,y): \dom  x  \rightarrow$ $W \times X \times Y$ for $\mathcal{H}$ is complete and satisfies the following: if $x(0,0) \in K$ then we have that $ x(t,j) \in K$ for all $(t,j) \in \dom x$.
	\end{itemize} 
\end{defn}

\begin{defn}
	Let $K \subseteq \mathbb{R}^{n}$ and $x \in K$, the contingent cone to set $K$ at point $x$, denoted $T_{K}(x)$, is given by:
	\begin{align}
		T_{K}(x)=\left\{z \in \mathbb{R}^{n} \mid \liminf _{h \to 0^{+}} \frac{d_{K}(x+h z)}{h}=0\right\}
	\end{align}
	where $d_{K}(y)$ denotes the distance of $y$ to $K$, defined by $d_{K}(y)=\inf _{y^{\prime} \in K}\left\|y-y^{\prime}\right\|$.
\end{defn}

In the sequel, we make the following assumption on the dynamics of the hybrid system \eqref{Eq2}.

\begin{assum}\label{assum1}
	The map $F: \mathbb{R}^n\times \mathbb{R}^m \rightrightarrows \mathbb{R}^n$ is outer semi-continuous, locally Lipschitz on $C$, locally bounded relative to $C \times  W$, and $F(x,w)$ is nonempty and convex for every $x \in C \times W$, and the output map $h$ is locally Lipschitz on $X$.
\end{assum}

Since it is possible to reason about cascade composition based on weak semantics, the analysis in this section will focus on feedback composition. Next, we define formally the feedback composition, consistent with Definition \ref{Def-feedbackComposition}, of the hybrid dynamical system \eqref{Eq2}.

\textbf{\underline{Hybrid systems under feedback composition}}

The feedback composition $\mathcal{H}_{f}$, derived from the hybrid system $\mathcal{H}=(W, X, Y, \mathcal{T})$ represented by \eqref{Eq2}, where $Y \subseteq W$, is formally defined as a hybrid dynamical inclusion expressed in the following form:
\begin{align}\label{Eq4}
	\mathcal{H}_{f}:\left\{\begin{aligned}
		x \in C_f & & \dot{x}  \in F_{f}(x) , \\
		x \in D_f	& & x^{+}\in G_{f}(x) ,
	\end{aligned}\right.
\end{align}
with $x(0,0)\in X^{0}$, $C_f=\{x \in X \mid (x,h(x))\in C\}$, $D_f=\{x \in X \mid (x,h(x)) \in D\}$,   $F_{f}=F(x,h(x))$ and $G_{f}=G(x,h(x))$.

\ifitsdraft
\else
\vspace{-5pt}
\fi

\subsection{Invariant relative to assume guarantee contracts}

%

Now, we introduce the notion of (pre-)invariant of hybrid dynamical systems relative to a contract.

\begin{defn}\label{def4.1}
	Let $\mathcal{H}=(W, X, Y, \mathcal{T})$ be a hybrid dynamical system described by \eqref{Eq2}. Let $\mathcal{C}=\left(A_{W}, G_{X}, G_{Y}\right)$ be an AG contract for $\mathcal{H}$ with a compact set of assumptions $A_{W}$. Consider a set $K \subseteq X$ such that $K \subset$ $\overline{C}_f \cup D_f$ and that $K \cap C_f$ is closed. Then, the set $K$ is said to be (pre-)invariant of $\mathcal{H}$ relative to the contract $\mathcal{C}$ if the following conditions hold:
	\begin{itemize}
		\item[(i)] $X^{0} \subseteq K \subseteq G_{X} \cap h^{-1}\left(G_{Y}\right)$;
		\item[(ii)] $G(K \cap D_f,A_W) \subseteq K$;
		\item[(iii)] For every $x  \in \widehat{C}_f , F\left(x, A_W\right) \subset T_{K \cap C_f}(x)$.
	\end{itemize}
	where the set-valued maps $F\left(., A_{W}\right)=\bigcup\limits_{w \in A_{W}} F(., w)$, $G\left(., A_{W}\right)=\bigcup\limits_{w \in A_{W}} G(., w)$ and $\widehat{C}_f=\partial (K\cap C_f)$. Furthermore, the set $K$ is said to be invariant of $\mathcal{H}$ relative to the contract $\mathcal{C}$ if every solution $(w,x,y): \dom  x  \rightarrow$ $W \times X \times Y$ is complete.
	
\end{defn}

\ifitsdraft
The essence of the above definition lies in the notion that if a hybrid dynamical system is forward (pre-)invariant within a set $K$, it implies that if the initial state $x(0,0)$ is in $K$, then any subsequent solution remains within the set $K$. Furthermore, if $K$ is a subset of $G_X\cap h^{-1}(G_Y)$, then every solution $x\in \mathcal{S}_{\mathcal{H}}(K)$ fulfils the specified contract $\mathcal{C}$. The following proposition shows that the existence of a (pre-)invariant for $\mathcal{H}$ relative to contract $\mathcal{C}$ implies the weak satisfaction of the contract $\mathcal{C}$.
\else
forward (pre-)invariant within a set $K$ implies that if the initial state $x(0,0)$ is in $K$, then any subsequent solution remains within the set $K$. Furthermore, if $K$ is a subset of $G_X\cap h^{-1}(G_Y)$, then every solution $x\in \mathcal{S}_{\mathcal{H}}(K)$ fulfils the specified contract $\mathcal{C}$. The following proposition shows that the existence of a (pre-)invariant for $\mathcal{H}$ relative to contract $\mathcal{C}$ implies the weak satisfaction of the contract $\mathcal{C}$.

\fi

\begin{prop}\label{prop2}
	Let $\mathcal{H}=(W, X, Y, \mathcal{T})$ be a hybrid dynamical system described by \eqref{Eq2} such that Assumption \ref{assum1} holds. Let $\mathcal{C}=\left(A_{W}, G_{X}, G_{Y}\right)$ be an AG contract for $\mathcal{H}$ with a compact set of assumptions. Then, $\mathcal{H} \models \mathcal{C}$, if there exists a closed set $K \subseteq X$ that is forward (pre-)invariant of $\mathcal{H}$ relative to the contract $\mathcal{C}$.
\end{prop}

\ifitsdraft
\begin{pf}
	It is sufficient to show the result for the (pre-)invariance. First, conditions (ii) and (iii) imply that conditions 1 and 2 in \cite[Theorem 4.3]{chai2018forward} are satisfied. Consider the map $\tilde{F}: \mathbb{R}^n \rightrightarrows \mathbb{R}^n$ defined for $x \in \mathbb{R}^n$ by $\tilde{F}(x)=F(x,A_W)$. Using the fact that $A_W$ is a compact set one has from Assumption \ref{assum1} that $\tilde{F}$ is outer semi-continuous, locally Lipschitz on $C$, locally bounded relative to $C$, and $\tilde{F}(x)$ is nonempty and convex for every $x \in C$. Hence, $K$, $C$, $D$ and $\tilde{F}$ satisfies \cite[Assumption 4.2]{chai2018forward}. It follows from \cite[Theorem 4.3]{chai2018forward} that for any hybrid arc $(x,y):\dom x  \to  X \times Y $ of the hybrid system
	\begin{align}
		\label{eq:6}
		\tilde{\mathcal{H}}:\left\{\begin{aligned}
			x \in C & & \dot{x} & \in F(x,A_W), \\
			x \in D & & x^{+}  & \in G(x,A_W) , \\
		   x(0,0) \in X^0 & &	y &= h(x),
		\end{aligned}\right.
	\end{align}
	initiated in the set $K$ (i.e $x(0,0) \in K)$ will satisfy $x(t,j)\in K$ for all $(t,j) \in \dom x$.
	Now we have all the ingredients to show that the system $\mathcal{H}$ in (\ref{Eq2}) weakly satisfies the contract $\mathcal{C}$. Consider a hybrid arc $(w,x,y):\dom x  \to W\times X \times Y $ of the hybrid system $\mathcal{H}$ in (\ref{Eq2}). Consider $(T, J) \in \dom  x$, and suppose that for all $(t, j) \in \dom  x \cap (\left[0, T\right]\times \left\lbrace0, \dots, J\right\rbrace)$, $w(t, j) \in A_{W}$. Since $x(0,0) \in X^0$ and $w(t,j) \in A_W$ for all $(t, j) \in \dom  x \cap (\left[0, T\right]\times \left\lbrace0, \dots, J\right\rbrace)$, one gets that the hybrid arc $(x,y):\dom  x \cap (\left[0, T\right]\times \left\lbrace0, \dots, J\right\rbrace)  \to  X \times Y $\footnote{The hybrid arc $(x,y): \dom x  \to  X \times Y $ can be obtained from the hybrid arc $(w,x,y):\dom x  \to W\times X \times Y$ by a simple projection over $X\times Y$.} is a hybrid arc of the hybrid system $\tilde{\mathcal{H}}$ in (\ref{eq:6}). Thanks to (i), from  $x(0,0) \in X^{0} \subseteq K$, one gets that for all $(t, j) \in \dom  x \cap (\left[0, T\right]\times \left\lbrace0, \dots, J\right\rbrace)$, $x(t,j) \in K \subseteq G_X$ and $y(t,j) \in h(K) \subseteq G_Y$. Hence, $\mathcal{H} \models \mathcal{C}$.  

\end{pf}
\fi

\ifitsdraft
\else
\vspace{-5pt}
\fi

\subsection{Invariant Composition}

We will now present results that enable reasoning about feedback composition based on the forward (pre-)invariance properties of hybrid dynamical systems.

\begin{thm}\label{THM5}(Invariants under feedback composition): Let $\mathcal{H}=(W, X, Y, \mathcal{T})$ be a hybrid dynamical system described by \eqref{Eq2} with maps and initial sets $F, G, h, X^{0}$, with $Y \subseteq W$ and let $\mathcal{H}_{f}=\left(X, \mathcal{T}_{f}\right)$. Let us assume that $F$ and $h$ are locally Lipschitz and Assumption \ref{assum1} holds for $\mathcal{H}$. Let $\mathcal{C}=\left(A_{W}, G_{X}, G_{Y}\right)$ be an AG contract for $\mathcal{H}$, with a compact set of assumptions $A_{W}$ and $G_{Y} \subseteq A_{W}$. If there exists a closed set $K \subseteq X$ (pre-)invariant of $\mathcal{H} $ relative to the contract $\mathcal{C}$, then, for all hybrid arcs $x: \dom x \to X$ in $\mathcal{T}_{f}$, we have $x(t,j) \in G_{X}$ for all $(t,j) \in \dom x$.
\end{thm}

\ifitsdraft
\begin{pf}
	It is sufficient to show the result for the (pre-)invariance. Let $x: \dom  x  \to X$ in $\mathcal{T}_{f}$. By definition of $\mathcal{H}_{f}$, $x: \dom  x  \to X$ is a solution to the following hybrid inclusion:
	\begin{align}
		\left\{\begin{aligned}
			\dot{x} & \in F_f(x)= F(x,h(x)) & & x \in C_f, \\
			x^{+} & \in G_f(x)= G(x,h(x)) & & x \in D_f,
		\end{aligned}\right.
	\end{align}
	
	Using the fact that $A_W$ is a compact set one has from Assumption \ref{assum1} that the map $F_f$ is outer semicontinuous, locally Lipschitz on $C_f$, locally bounded relative to $C_f$, and $F_f(x)$ is nonempty and convex for every $x \in C_f$. Hence, $K$, $C_f$, $D_f$ and $F_f$ satisfies \cite[Assumption 4.2]{chai2018forward}.
	
	Moreover, we have that $G_f(K\cap D_f)=G(K\cap D_f, h(K\cap D_f)) \subseteq G(K \cap D_f,h(K)) \subseteq G(K \cap D_f,G_Y) \subseteq G_f(K \cap D_f, A_W)$, where the second inclusion comes from (i) in Definition \ref{def4.1}, and the last inclusion from the fact that $G_Y \subseteq A_W$. Hence, conditions 1 in \cite[Theorem 4.3]{chai2018forward} is satisfied. Now consider $x \in \widehat{C}_f=\partial(K\cap C_f)$, we have that $F_f(x)=F(x,h(x))\subseteq F(x,h(K\cap C_f)) \subseteq F(x,h(K)) \subseteq F(x,G_Y) \subseteq F(x,A_w)$, where the first inclusion follows from the fact that $K \cap C_f$ is closed, the third inclusion comes from (i) in Definition \ref{def4.1}, and the last inclusion from the fact that $G_Y \subseteq A_W$. Hence, conditions 2 in \cite[Theorem 4.3]{chai2018forward} is satisfied. Since $x(0,0) \in X^0 \subseteq K$, one has from \cite[Theorem 4.3]{chai2018forward} that for all $(t,j) \in \dom x$, $x(t,j) \in K \subseteq G_X$, which ends the proof.
	
\end{pf}
\fi

\ifitsdraft
We present an illustrative example to demonstrate the practical application of the preceding theorem.

\begin{exmp} \label{Example4}
	Consider the following hybrid system given by,
	$$
	\mathcal{H}_{3} : 
	\begin{cases}
		x \in C & \;\dot{x} \;=F(x,w):=-2 x-2 w, \\
		x \in D & x^{+}=G(x):=\frac{x}{2},  \\
		& y\;\;=h(x) \; :=x,
	\end{cases}
	$$
	where the flow sets are $C=\left\{x \in \mathbb{R}: |x| \leqslant b\right\}$, and the jump sets are $D=\left\{x \in \mathbb{R}\mid \right. $ $\left. x = 0\right\}$,  $b \in \mathbb{R}_{>0}$. The set of initial condition $X^0=\left\{x \in \mathbb{R}: |x| \leqslant b\right\}$. We observe that during flow, solutions tend continuously to the origin; while at jumps, solutions remain at the current location (the origin). 
	Consider the AG contract $\mathcal{C}=([-b,b],[-b,b],[-b,b])$ and the set $K=\left[-b,b\right]$, one can easily check that $X^{0} \subseteq K \subseteq G_{X} \cap h^{-1}\left(G_{Y}\right)$, $G(K \cap D,A_W) \subseteq K$ and for every $x  \in \partial C , F\left(x, A_W\right) \subset T_{K \cap C}(x)$, since
	$$
	T_{[-b, b]}(x)=\left\{\begin{array}{lll}
		\mathbb{R}_{\geqslant0} & \text {if } & x=-b, \\
		\mathbb{R}^{-} & \text {if } & x=b, \\
		\mathbb{R} & \text { if } & x \in(-b, b)
	\end{array}\right.
	$$
	Moreover, we have that $K \subset$ $\overline{C} \cup D$ and that $K \cap C$ is closed. Hence, all conditions of Definition \ref{def4.1} are satisfied and $K$ is an invariant of $\mathcal{H}_3$ relative to the contract $\mathcal{C}$. Moreover, one can easily check that Assumption \ref{assum1} is satisfied for the hybrid system $\mathcal{H}_3$. Let $\left(\mathcal{H}\right)_{f}=\left(X, \mathcal{T}_{f}\right)$, the hybrid arcs $x: \dom  x \to X$ in $\mathcal{T}_{f}$ are the solutions of the hybrid system: 
	$$
	\left(\mathcal{H}\right)_{f} \begin{cases}
		\dot{x}=-4x & x \in C_f,\\
		x^{+}=\frac{x}{2} & x \in D_f,
	\end{cases}
	$$
	with $x(0,0) \in\left[-b, b\right]$. By Theorem \ref{THM5}, it follows that for all $(t,j) \in \dom  x, x(t,j) \in[-b, b]$.
\end{exmp}

\begin{figure}
	\centering
	\includegraphics[width=1\linewidth]{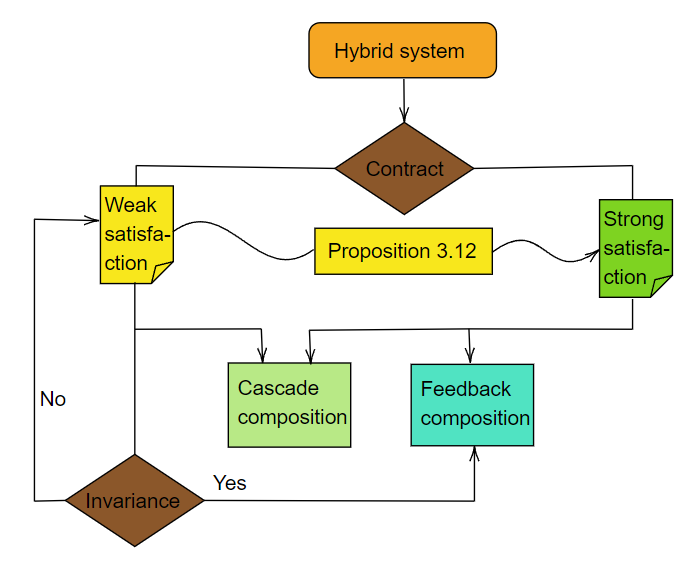}
	\caption{Overview of Key Findings}
	\label{fig:conclusion}
\end{figure}
\fi

\section{Conclusion}
\ifitsdraft
\else
\vspace{-10pt}
\fi

\ifitsdraft
This work verifies contract-based designs in interconnected hybrid systems where compliance may shift over time, impacting compositions. In particular, we define an AG contract for hybrid dynamical systems. Results show that both weak and strong semantics are compatible with cascade composition. In light, it is shown that strong semantics are necessary for the analysis of feedback composition (Theorem \ref{THM2}). Furthermore, under certain assumptions, it is possible to extend weak contract satisfaction to strong satisfaction (Proposition \ref{prop1}). 	New conditions under which forward (pre-)invariance of a hybrid system implies weak contract satisfaction are established (Proposition \ref{prop2}). The forward (pre-)invariants are compatible with feedback compositions (Theorem \ref{THM5}). The main results of this work are summarized in Figure \ref{fig:conclusion}, which illustrates how to reason about compositions of contracts. In our future work, we will extend our framework to handle properties such as linear temporal logic and develop tools for verifying contract satisfaction by a hybrid system or constructing controllers to enforce its satisfaction.
\else
This study analyses contract-based designs in interconnected hybrid systems, where compliance may shift over time, affecting compositions. We define an assume-guarantee contract for hybrid dynamical systems and find both weak and strong semantics compatible with cascade composition. However, strong semantics are essential for feedback composition analysis (Theorem \ref{THM2}). Additionally, we explore conditions allowing the extension of weak contract satisfaction to strong satisfaction (Proposition \ref{prop1}). Furthermore, new conditions are established where forward (pre-)invariance of a hybrid system implies weak contract satisfaction (Proposition \ref{prop2}), which is compatible with feedback compositions (Theorem \ref{THM5}). In our future work, we will extend our framework to handle properties such as linear temporal logic and develop tools for verifying contract satisfaction by a hybrid system or constructing controllers to enforce its satisfaction.
\fi


\bibliography{ifacconf}             

\appendix
\end{document}